\documentclass[twocolumn,superscriptaddress,showpacs,amsmath,amssymb,aps]{revtex4}

\usepackage{graphicx}
\usepackage{dcolumn}
\usepackage{bm}
\usepackage{units}

\newcommand{\grad}{\ensuremath{^{\circ}}}
\newcommand{\STO}{$\text{SrTiO}_{3}$}

\begin{document}

\title{Metal-insulator transition on SrTiO$_{3}$ surface induced by ionic-bombardment}

\author{Heiko Gross}
\affiliation{Physikalisches Institut der Universit\"at W\"urzburg, Am Hubland, 97074 W\"urzburg, Germany}
\affiliation{Department of Physics and Astronomy, Rutgers, The State University of New Jersey, 136 Frelinghuysen Road, Piscataway, NJ 08854-8019, USA}
\author{Namrata Bansal}
\affiliation{Department of Electrical and Computer Engineering, Rutgers, The State University of New Jersey, 94 Brett Road, Piscataway, NJ, USA}
\author{Yong-Seung Kim}
\author{Seongshik Oh}
\email{ohsean@physics.rutgers.edu}
\affiliation{Department of Physics and Astronomy, Rutgers, The State University of New Jersey, 136 Frelinghuysen Road, Piscataway, NJ 08854-8019, USA}

\date{\today}

\begin{abstract}
\STO\ (STO) is one of the most popular insulating single-crystal substrates for various complex-oxide thin film growths, because of its good lattice match with many complex oxide films. Here, we show that a common thin film processing technique, argon ion-milling, creates highly conducting layer on the surface of STO, not only at room temperatures but also at cryogenic temperatures at which thermal diffusion is completely suppressed. Systematic \emph{in situ} four-point conductance measurements were taken on single-crystal STO substrates inside vacuum environment. The evolution of metallicity out of insulating STO follows simple models based on oxygen vacancy doping effect. At cryogenic temperatures, ion milling created a thin - but much thicker than the argon-penetration depth - steady-state oxygen-vacant layer, leading to a highly-concentric metallic state. Near room temperatures, however, significant thermal diffusion occurred and the metallic state continuously diffused into the bulk, leaving only low concentraion of electron carriers on the surface. Analysis of the discrepancy between the experiments and the models also provided evidence for vacany clustering, which seems to occur during any vacancy formation process and affects the observed conductance. These observations suggest that the transport properties of films processed on STO substrates using energetic methods such as ion milling need to be taken with caution. On the other hand, if properly controlled, ionic bombardment could be used as a way to create selective conducting layers on the surface of STO for device applications.

\end{abstract}

\maketitle

\section{Introduction}
\label{intro}
Strontium titanate is one of the most studied perovskite-like oxides, which exhibit a variety of features such as high $\text{T}_{\text{c}}$ superconductivity, colossal magnetoresistance and ferroelectricity \cite{Hwang:08,Cava:87,ramirez:97,Cohen:93}. Stoichiometric \STO\ has an indirect band gap of $\unit[3.25]{\text{eV}}$ and is therefore a band insulator \cite{benthem:6156}. However, its electronic property can be changed by doping \cite{luo:04,astala:01}; cation-substitutions like La for Sr or Nb for Ti \cite{tokura:93,Shanthi:98,astala:02} or oxygen vacancies \cite{luo:04,Ricci:03, Szot:02, Szot:06, Meijer:05} introduce n-type carriers, making STO metallic. Doping levels that create a carrier concentration as low as $\unit[2.7\times10^{17}]{\text{cm}^{-3}}$ are sufficient to transform insulating STO into a metallic state \cite{Tufte:67}. For certain doping levels, even superconductivity with $\text{T}_{\text{c}}<\unit[0.3]{\text{K}}$ has been reported \cite{Schooley:64}. In particular, oxygen vacancies contribute more than just electric carriers. They exhibit, if in great amounts, dark-blue fluorescence through a transition between their impurity level and the valence band \cite{kan:07}. Oxygen vacancies are also actively investigated for resistive memory applications \cite{Szot:06,sawa:06,beck:00}. 

In recent years, researchers found that when thin films are grown on STO substrate by pulsed laser deposition (PLD), energetic species during the growth process induce oxygen vacancies and create a metallic layer into the substrate \cite{Hwang:04}. Similarly, ion-milling process, a common dry-etching technique, was also found to introduce metallic states on insulating STO by creating oxygen vacancies \cite{kan:07,reagor:04,kan:05}. These observations raise serious questions on transport measurements for films processed on STO substrates. Considering the routine usage of energetic processes such as PLD and ion milling with STO substrates, proper understanding of how energetic particles create conducting states on STO is important. In order to investigate how metallic states emerge on STO surface through energetic bombardment, it is essential to carry out the experiment without exposing samples to air because oxygen in air can quickly fill the vacancies on the surface and mask metallic states underneath. Such studies do not exist yet. Here, by combining ion-milling process, \emph{in situ} conductance measurement and model calculations, we show that metallic states emerge even when thermal diffusion is completely suppressed at cryogenic temperatures, and also show how such conducting states emerge and evolve during and after ion milling in different conditions.

\section{Experimental setup}
\label{experiment}
The experiment was carried out in a custom-designed high vacuum chamber (base pressure less than $\unit[10^{-7}]{\text{Torr}}$) that is equipped with an \emph{in situ} four point probe having springy contacts; see Ref.\cite{gross:09} for the details of the setup. This unique system allows Ar-ion bombardment over a wide range of temperatures ($\unit[-160]{\grad C}$ to $\unit[700]{\grad C}$) and four-point-probe conductance measurements within the same vacuum condition. We used a gridlness ion source (EH200F, Kaufman \& Robinson), which allows large ion beam currents even at very low beam energies down to $\unit[50]{\text{eV}}$. We also measured oxygen partial pressures using a residual gas analyzer (RGA), and at the base pressure, the oxygen partial pressure was found to be less than $\unit[10^{-9}]{\text{Torr}}$. For each experiment, a polished single-crystal (001) STO substrate of $\unit[10]{\text{mm}} \times\unit[10]{\text{mm}} \times \unit[0.5]{\text{mm}}$ (MTI and Crystec \emph{as received}) was mounted with silver paint on a copper stage, which was then cooled with either liquid nitrogen or water depending on the target temperatures. During ion-milling process, the four point probe was positioned far away from the sample so that it does not interfere with the ion beam. After each ion-milling process, contact between the four point probe and the sample was made using a manual linear-motion feedthrough, and the conductance was measured using a source meter (Keithley 2636A). Over the entire process, temperature of the sample was monitored using a thermocouple mounted on a corner of the substrate.  

\section{Results}
\subsection{Model at cryogenic (100 $\sim$ 200K) temperatures}
\label{cryo}
Because each non-clustered oxygen vacancy donates two electrons and the electronic mobility is fairly constant over a wide range of doping level at temperatures above 100 K in STO \cite{calvani:93}, to the first approximation we can assume that the electric conductivity of STO measured by the four point probe is proportional to the density of oxygen vacancies. With this assumption, we first construct a simple model on how oxygen vacancies form during ion-milling process when thermal diffusion is neglected and will use it to describe the conductance measurements at cryogenic temperatures in the following section. 

\begin{figure}
\includegraphics[width=1.0\linewidth]{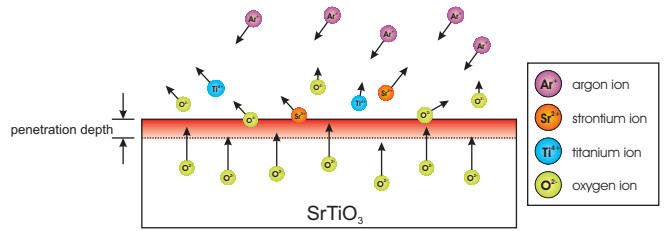}
\caption[Scheme of ion bombardment]{(Color online) Development of an oxygen deficient layer beyond the penetration depth of argon ions. The surface layer is continuously removed during argon-ion milling process. Oxygen vacancies tend to diffuse deeply into STO surface because of the significantly higher diffusivity of oxygen than those of strontium and titanium.}
\label{scheme_bombardment}
\end{figure}

Oxygen is not the only type of ion that is removed during ion milling of the STO surface. In fact, the whole surface gets removed, which gives rise to a question why oxygen vacancies are formed at all \cite{kan:07,reagor:04}. Formation of oxygen vacancies can be understood by considering the diffusivity difference between ions; the diffusivity of oxygen is significantly higher than those of titanium and strontium for multiple reasons. First of all, oxygen is lighter and is thus easier to move in the lattice. Second, its nearest neighbor distance is smaller. The distances of Ti and Sr are by a factor of $\sqrt{2}$ larger than that of oxygen. Furthermore, oxygen has a free line of sight to the nearest neighbor, which is not the case at least for titanium. Because diffusivity depends almost exponentially on these parameters, their combined effect leads to a significant difference in the diffusivities of oxygen vs. strontium and titanium. Higher diffusivity of the oxygen ion results in the formation of a oxygen vacant layer deeper than argon can penetrate into STO. This scenario is depicted in Fig. \ref{scheme_bombardment}, where argon-ion penetration depth is smaller than $\unit[3]{\text{nm}}$ for beam energies below $\unit[300]{\text{eV}}$ \cite{reagor:04, kan:05}. As the top surface layer is being etched away by incoming argon ions, the oxygen ions below that layer are able to fill out the vacancies within the penetration depth by thermal migration. Through this process, oxygen vacancies can be formed in a region much deeper than the penetration depth of argon ions \cite{kan:05}. 

Considering that the diffusivity is exponentially dependent on temperature, the thermal state of the \STO\ sample during ion-milling must be a critical factor for the depth profile of the oxygen vacancies. Based on the diffusion coefficients available in the literature for oxygen vacancies \cite{Pasierb:99}, at temperatures below $\unit[-100]{\grad\text{C}}$, the diffusion rate for oxygen vacancies is estimated to be much smaller than the ion milling rate. This implies that thermal diffusion can be neglected at these temperatures because incoming argon ions will etch away oxygen vacancies before they diffuse into the bulk. If such is the case, from Fig. \ref{scheme_bombardment} the oxygen deficient layer should be limited to the argon penetration depth, and the density of oxygen vacancies must reach a steady-state value almost immediately after ion-milling process is started. However, it was not consistent with our experimental observation described later, which strongly suggests that the oxygen deficient layer extends much beyond the argon penetration depth even at these cryogenic temperatures. We propose that the lattice vibrations on the surface, caused by collisions with the argon ions, are the reason for the oxygen vacant layer developing beyond the penetration depth of argon ions. The intensity of the vibrations decays exponentially with depth, and the thickness of the vibrating layer, which we call \emph{hot-zone}, is determined by the kinetic energy of the incoming argon ions. Because the incoming argon ions are incoherent, coherent vibration modes, i.e. phonons, cannot propagate through the bulk. The \emph{hot-zone} formation is also different from the standard heating process in that it cannot propagate into the crystal beyond the vibration decay length. 

\begin{figure}[ht!]
  \begin{center}
    \includegraphics[width=1.0\linewidth]{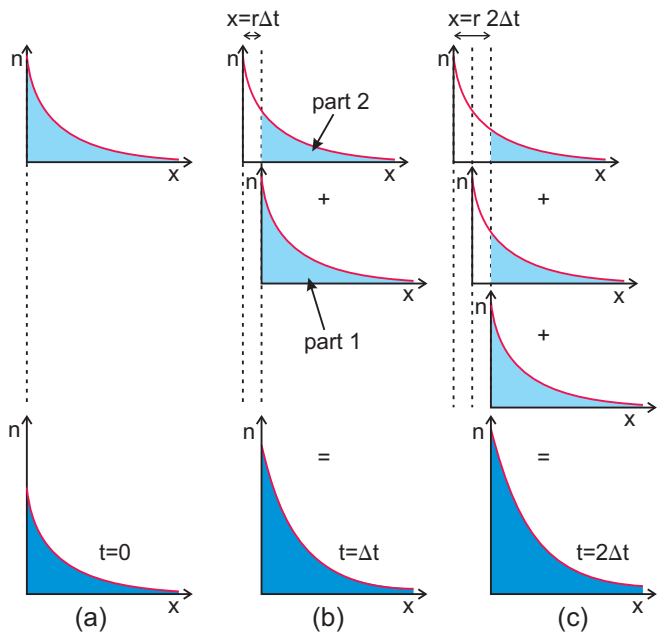}
  \end{center}
    \caption[Development of \textit{steady-state}]{(Color online) Formation of the oxygen vacancy layer during ion-milling process. (a) At time $t=0$, the oxygen vacancy distribution forms an exponential shape following the decay of the vibration energy in the \textit{hot-zone}. (b) After a time interval $\Delta t$, the ion-milling process removes a layer with thickness $r\cdot \Delta t$, where $r$ is the milling rate. So the vacancy concentration profile moves about this depth further in addition to the remaining part of the previous profile. (c) As this process continues, the original profile ($t=0$) gets gradually removed.}
    \label{scheme_hot-zone}
\end{figure}

Considering that the propagation rate of oxygen vacancies through the \emph{hot-zone} must be much higher than the ion-milling rate, during short time $\Delta t$, we can assume that the oxygen vacancies propagate through the \emph{hot-zone} almost immediately before any significant ion milling occurs. If we assume an exponentially decaying depth profile of the oxygen vacancies during this short time, the depth profile of the carrier concentration, which is assumed to be proportional to that of the oxygen vacancies, should increase by
\begin{equation}
\Delta n\left(x,\Delta t\right)=\frac{dn_{\text{s}}}{dt}\:\Delta t \:e^{-\frac{x}{d_{0}}},
\label{carrierProfile}
\end{equation}
where $\frac{dn_{\text{s}}}{dt}$ is the rate of concentration increase at the very surface and $d_{0}$ is the effective \emph{hot-zone} thickness. Additional factor to consider is the removal of the top-most layer during this time. With the milling rate $r$, the thickness of the removed layer should be $\Delta x=r\cdot\Delta t$ after a time interval $\Delta t$. After each $\Delta t$, new carriers are created on the surface (part 2) with the same profile as in Eq. (\ref{carrierProfile}) and added to the already existing carriers (part 1) as shown in Fig. \ref{scheme_hot-zone}.
If we integrate over the whole depth taking $\Delta t$ infinitesimally small, this model leads to a areal carrier concentration of 
\begin{equation}
N\left(t\right)=\frac{dn_{s}}{dt}\,\frac{d_{0}^{2}}{r}\left(1-e^{-\frac{r}{d_{0}}t}\right).
\label{carrierEquation}
\end{equation}
Here, the exponential term vanishes after long time and the concentration saturates to a \emph{steady-state} value; note that the saturation will occur faster as the milling rate $r$ increases. If there were not any etching process (that is, $r = 0$), the areal carrier concentration introduced per time, $\frac{dN_{no-etching}}{dt}$ would reduce to $\frac{dn_{s}}{dt}d_{0}$, which implies that the higher the beam current is (larger $\frac{dn_{s}}{dt}$) or the higher the ion energy is (larger $d_{0}$), the more vacancies are created. It is reasonable to assume that this value should be proportional to the milling rate $r$, because milling is nothing more than vacancy formation of all the component ions. In other words, 
\begin{equation}
\frac{dN_{no-etching}}{dt}=\frac{dn_{s}}{dt}d_{0}\propto r.
\end{equation}
Putting this expression into Eq. (\ref{carrierEquation}) leads to the simplified equation
\begin{equation}
N\left(t\right)\propto d_{0}\left(1-e^{-\frac{r}{d_{0}}t}\right).
\label{saturation}
\end{equation}
This result provides an important insight: the saturation carrier density (as t $\rightarrow\infty$) is proportional to $d_{0}$, which should grow with the argon energy just as the argon penetration depth increases with the beam energy \cite{harper:82}. So we expect that higher beam energy will lead to higher saturation value of the areal carrier density on ion-milled STO surface at cryogenic temperatures.

\subsection{Measurement at cryogenic temperatures}
\begin{figure}[ht!]
  \begin{center}
    \includegraphics[width=1.0\linewidth]{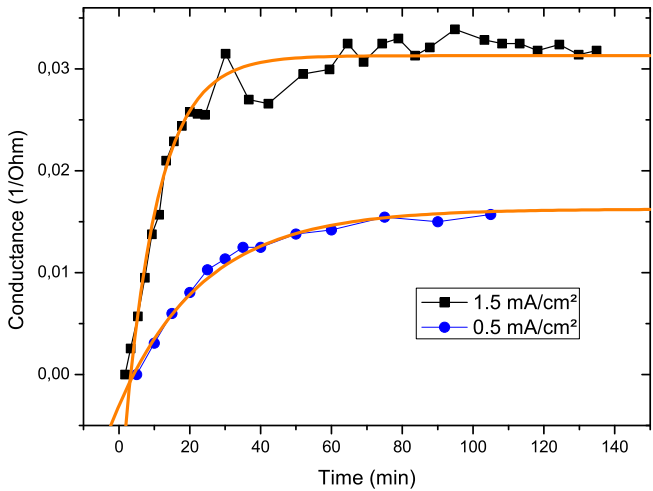}
  \end{center}
    \caption[Saturation of conductance for different beam currents]{(Color online) Conductance of \STO\ versus ion-milling time with a beam energy of $\unit[66]{\text{eV}}$ and different ion beam currents. The ion-milling temperature never exceeded $\unit[-100]{\grad\text{C}}$, and the conductance measurement was performed at $\unit[-140]{\grad C}$. The fitting curves are based on Eq. (\ref{carrierEquation}).}
    \label{lowTempDifferentCurrent}
\end{figure}

Figure \ref{lowTempDifferentCurrent} shows how electric conductance of fresh insulating \STO\ substrates changes as a function of ion-milling time with the sample at cryogenic temperatures (below $\unit[-100]{\grad\text{C}})$, for beam energy of $\unit[66]{\text{eV}}$ and two different beam current densities. 
The saturation of the conductance after about an hour is consistent with the cryogenic-temperature model, Eq. (\ref{carrierEquation}). According to the model, the thickness of the conducting \emph{hot-zone}, $d_{0}$, must be the same for the two curves in Fig. \ref{lowTempDifferentCurrent}, because it depends only on the beam energy.  The fitting parameter for the exponent $\frac{r}{d_{0}}$ of the $\unit[1.5]{\text{mA}/\text{cm}^2}$ curve is 2.5 times higher than that for the $\unit[0.5]{\text{mA}/\text{cm}^2}$ curve. Despite the simplicity of our model, this value is in reasonable agreement with the factor of three ratio of the beam currents, which is also proportional to the ion-milling rate, $r$. 

However, the fact that the saturation values of the two curves are significantly different is inconsistent with Eq. (\ref{saturation}), according to which the same saturation value is expected for both curves because they have the same beam energy. We propose that this discrepancy is partly due to clustering of oxygen vacancies; this clustering idea is further supported by follow-up measurements discussed below. When oxygen vacancies form clusters, they start to trap electrons, lowering the conductivity from the value predicted from non-clustered vacancies \cite{gong:91}. Because clustering is energetically favored \cite{Cuong:07}, unless vacancy diffusion is completely forbidden, it should be impossible to avoid the formation of clusters. Even at cryogenic temperatures as in Fig. \ref{lowTempDifferentCurrent}, although thermal diffusion beyond the \emph{hot-zone} is suppressed, diffusion will still occur within the \emph{hot-zone} during ion milling, and thus nearby vacancies can form clusters. The longer vacancies stay within the \emph{hot-zone} during ion milling, the more clusters will form, resulting in lower electrical conductivity. This scenario is consistent with Fig. \ref{lowTempDifferentCurrent}. The crucial factor is the time it takes to refresh the \emph{hot-zone}. This time can be calculated out of the milling rate $r$ and the thickness $d_{0}$. The faster milling rate of the $\unit[1.5]{\text{mA}/\text{cm}^2}$ measurement leads to a higher refreshing rate, so the oxygen vacancies have less time to cluster and the electric conductance should be higher than that of the $\unit[0.5]{\text{mA}/\text{cm}^2}$ measurement.

The thickness of the conducting layer, say for the $\unit[1.5]{\text{mA}/\text{cm}^2}$ curve, is estimated as $d_{0}=\unit[57]{\text{nm}}$ from the fitting parameter and measured milling rate. It should be noted that this thickness is that of the oxygen deficient layer and is much larger than the argon penetration depth, which is only a few nanometers \cite{reagor:04}. Based on the effective thickness of the conducting layer, $d_{0}$, and the electron mobility ($\mu=\unit[49]{\frac{\text{cm}^{2}}{\text{Vs}}}$) of conducting STO at the measurement temperature ($\unit[-140]{\grad C}$) \cite{Frederikse:67}, the \emph{steady-state} surface carrier density,  $n_{s}(=\frac{N(t\rightarrow\infty)}{d_{0}})$, is estimated as $n_{s,66eV}=\unit[2.0\times10^{20}]{\text{cm}^{-3}}$, which is comparable to high end values due to oxygen vacancies \cite{Frederikse:67}. This is equivalent to about $0.3\%$ oxygen vacancies at the top surface, if every oxygen donates two electrons. 

\begin{figure}[ht!]
  \begin{center}
    \includegraphics[width=1.0\linewidth]{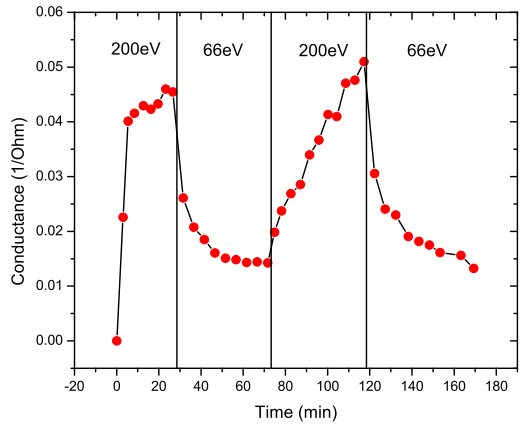}
  \end{center}
    \caption[Switching conduction in STO due to changing beam energies at low temperature]{(Color online) Conductance switching of \STO\ at $\unit[-130]{\grad C}$ with alternating ion beam energy and constant beam current density of $\unit[0.5]{\text{mA}/\text{cm}^2}$. Higher energy provides a thicker \textit{hot-zone}, and thus higher conductance. Switching back to $\unit[66]{\text{eV}}$ causes the conductance to decrease as the conducting layer is getting thinner. Such switching behavior did not occur near room temperature.}
    \label{conductanceSwitching}
\end{figure}

According to our model, if the beam energy is changed, thickness of the \emph{hot-zone}, accordingly the conductance, should also change; higher beam energies should result in thicker oxygen deficient layer, thus higher conductance. This is seen in Fig. \ref{conductanceSwitching}, which shows that the conductance of the sample can be changed back and forth by switching the beam energies.
The conductance depends on the beam energy because the thickness of the \emph{hot-zone} changes. It is noted that the observed behavior is a combination of various effects. The influence of clusters and increasing surface roughness are factors that complicate the development of an exact model, but it is intriguing that our simple model can still explain the main features of the observed phenomena.

\subsection{Model near room temperatue}
\label{roommodel}
Measurements near room temperature showed a totally different behavior. It was found that the thermal diffusion of oxygen vacancies can be described by Fick's second law \cite{Pasierb:99}. The significantly enhanced mobility of vacancies at room temperature compared to that at cryogenic temperatures leads to much thicker conducting layers with reduced carrier densities. Nevertheless, it is still reasonable to assume that a \emph{steady-state} carrier concentration, $n_{s}$, should be reached at the top surface at a certain time after ion milling initiates; this will be the case when the amount of vacancies migrating from the surface into the bulk equilibrates with the amount of vacancies introduced onto the surface by ion-milling. This constraint of a constant surface concentration leads to the following time and space dependent solution of Fick's second law
\begin{equation}
n\left(x,t\right)=n_{s}\:\text{erfc}\left(\frac{x}{2D\sqrt{t}}\right),
\end{equation}
where $D$ is the vacancy diffusion constant, $x$ is the depth from the surface, and erfc is the complementary error function. This equation describes development of the carrier density per volume with increasing depth $x$ and a constant surface concentration $n_{s}$. Integrating this function over the entire thickness leads to the following areal carrier concentration: 
\begin{equation}
N_{\text{total}}\left(t\right)=\frac{2n_{s}\sqrt{D}}{\sqrt{\pi}}\sqrt{t}.
\label{surfaceConcentrationEquation}
\end{equation}
It is noted that this areal carrier concentration never saturates, unlike the cryogenic case above; this implies that the oxygen vacancies contiously diffuse into the bulk at these elevated temperatures. 

\subsection{Measurement near room temperatue}
\label{roommeasure}
\begin{figure}[ht!]
  \begin{center}
    \includegraphics[width=0.9\linewidth]{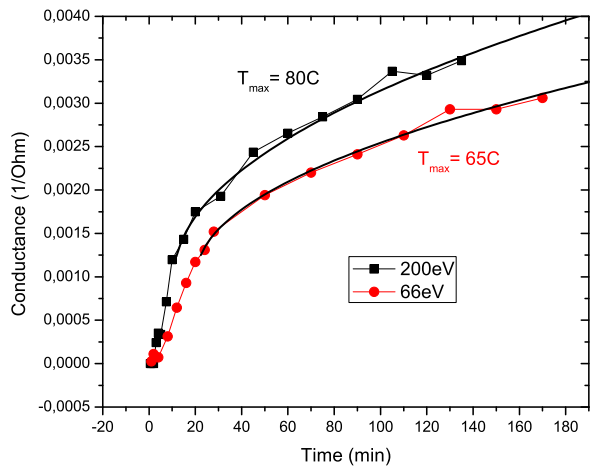}
  \end{center}
    \caption[Comparison of different beam energies]{(Color online) Conductance versus milling time for two different beam energies near room temperature. The behavior is completely dominated by thermal diffusion. The fit is done with Eq. \ref{surfaceConcentrationEquation}. The slightly higher conductance of the $\unit[200]{\text{eV}}$ curve is accounted for by the higher sample temperature during ion milling induced by its higher beam energy. The conductance measurement was done at $\unit[30]{\grad C}$.} 
    \label{Conductance66eV200eV}
\end{figure}

The experimental results are consistent with this diffusion model as shown in Fig. \ref{Conductance66eV200eV}.
The plot shows the development of the conductance with different beam energies and with the same beam current. The fit with Eq. \ref{surfaceConcentrationEquation} is valid only after a certain time because it takes time to reach the \emph{steady-state} surface concentration, which is the assumption used for Eq. \ref{surfaceConcentrationEquation}. The effective thickness of the oxygen deficient layer increases over time with $\sqrt{t}$, and, at $t = 100 \text{min}$, is estimated as $d_{\text{eff},\text{66eV}}=\unit[2.4]{\mu m}$ for the $\unit[66]{\text{eV}}$ case, and $d_{\text{eff},\text{200eV}}=\unit[3.7]{\mu m}$ for the $\unit[200]{\text{eV}}$. The estimated surface concentrations are $n_{s,\unit[66]{\text{eV}}}=\unit[1.8\times10^{18}]{\text{cm}^{-3}}$ and $n_{s,\unit[200]{\text{eV}}}=\unit[1.5\times10^{18}]{\text{cm}^{-3}}$ for the $\unit[66] {\text{eV}}$ and the $\unit[200] {\text{eV}}$ case, respectively, which are more than two orders of magnitude smaller than the value, $n_{s,66eV}=\unit[2.0\times10^{20}]{\text{cm}^{-3}}$, found above at cryogenic temperatures. These numbers are all physically quite reasonable and suggest that both our low temperature model with \emph{hot zone} and high temperature model with thermal diffusion explain the main mechanisms behind ion-milling induced metal-insulator transitions on STO.

\subsection{Thermal cycle and Time dependence}
\label{cycle}
\begin{figure}[ht!]
  \begin{center}
    \includegraphics[width=1.0\linewidth]{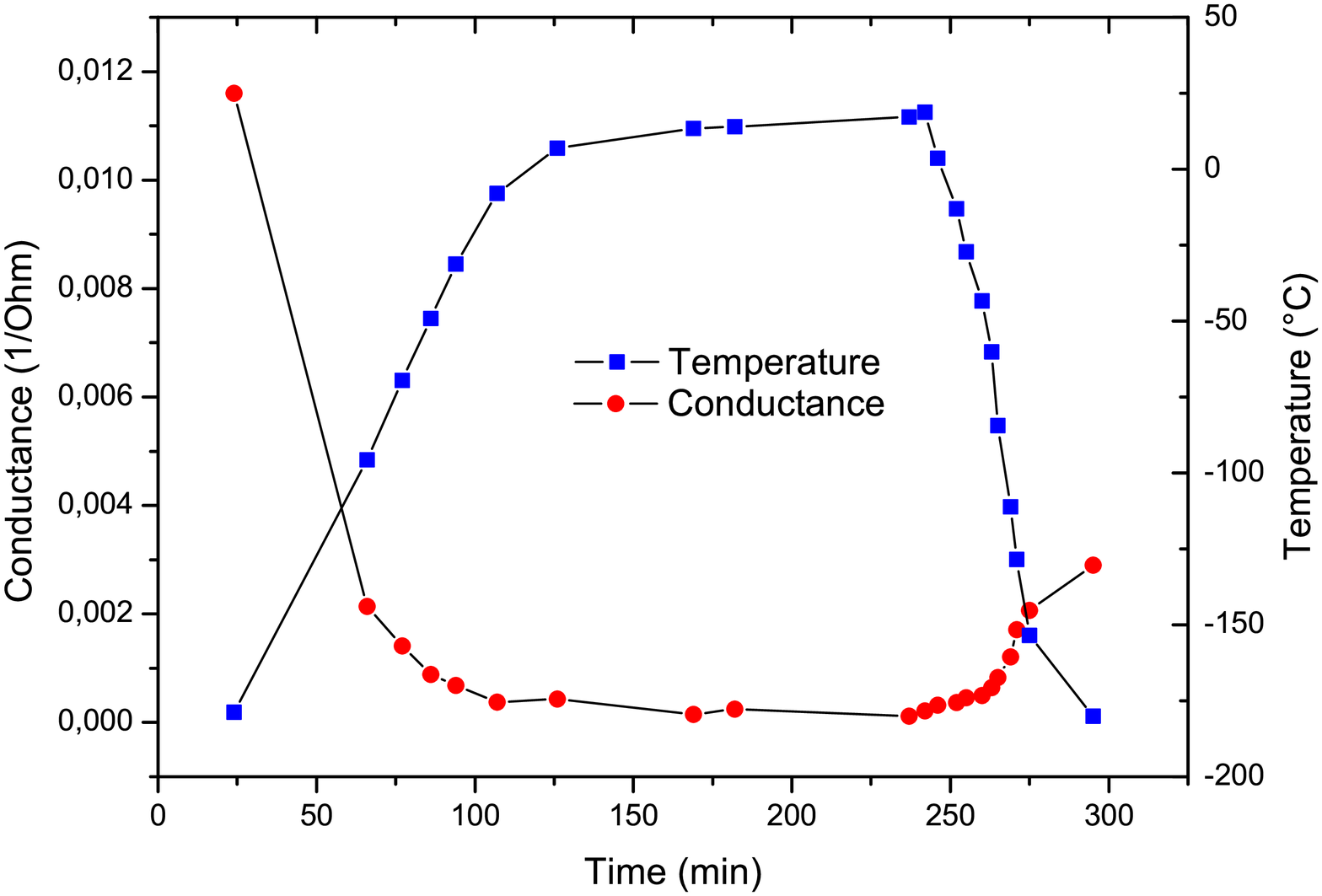}
  \end{center}
    \caption[Influence of oxygen vacancy clustering on the conductance of \STO]{(Color online) Conductance change due to thermal cycle up to room temperature. After ion-milling at cryogenic temperatures, the sample conductance was contiually measured during the warm-up and cool-down. Note that the low temperature conductance significantly (factor of four) dropped after the thermal cycle. In contrast, maintaining the sample at cryogenic temperatures preserved the conductance for over 20 hours.}
    \label{ClusterFactorGraph}
\end{figure}

Figure \ref{ClusterFactorGraph} shows how the conductance of a cryogenically ion-milled sample changes over a thermal cycle up to room temperature. It shows that an irreversible change of carrier concentration occurred during the cycle, and this phenomena cannot be explained by our simple models assuming two electrons per oxygen vacancy. We propose that vacancy clustering should be the most likely mechanism behind this effect. 
In Fig. \ref{ClusterFactorGraph}, a fresh insulating STO was ion-milled at a cryogenic temperature (below $\unit[-100]{\grad C}$) and its conductance was measured through a thermal cycle from $\unit[-180]{\grad C}$ up to room temperature.  As the sample is warmed to room temperature, its conductance should drop because the electron mobility decreases in STO as temperature rises \cite{Frederikse:67}. If this is the only mechanism affecting the conductance, the initial conductance should be recovered when the sample is cooled back to its initial temperature. Instead, the conductance dropped by a factor of four. 

There are two mechanisms that can explain this irreversible process: one is oxygen vacancy clustering and the other is surface oxidation. Both are diffusion-limited, and thus suppressed at cryogenic temperatures, but as the sample warms up, both mechanisms become active and cause the conductance to decrease. Considering that the entire experiment was performed in a high vacuum chamber with oxygen partial pressure less than $\unit[10^{-9}]{\text{Torr}}$, the oxidation effect is less likely and thus clustering seems to be the dominant effect causing this phenomenon. Thermal diffusion, which is negligible at cryogenic temperatures, enables oxygen vacancies to migrate and form clusters near room temperature. This state cannot be reversed upon re-cooling because clusters are energetically favorable \cite{Cuong:07,Cordero:09,Cherry:95}. When two vacancies join to form a cluster, only half of the original carriers remain mobile due to the trapping mechanism \cite{Cuong:07}. Accordingly, clustering with two vacancies per cluster would yield a conductance drop of only 50\% even if the entire vacancies have clustered. The observed factor-of-four conductance drop, therefore, implies that significant portion of the clusters were composed of more than two vacancies per cluster. 

\begin{figure}[ht!]
  \begin{center}
    \includegraphics[width=1.0\linewidth]{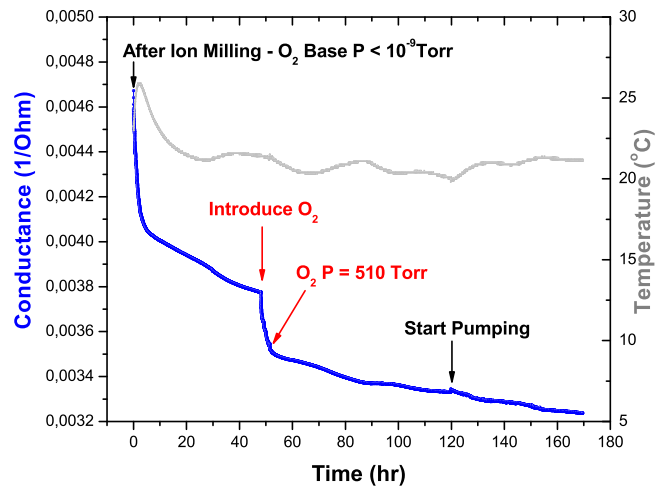}
  \end{center}
    \caption[Oxidation and clustering of oxygen vacancies over time]{(Color online) Conductance decrease of an ion-milled STO surface over a long period of time with exposure to increasing oxygen pressure. Both the ion-milling and the measurement were carried out at room temperature. Note that the gradual conductance decrease before and after 50 hours and after 120 hours remain almost unaffected even with orders of magnitude change in the oxygen pressure.}
    \label{oxidationSTO}
\end{figure}

Even though the partial oxygen pressure during the measurement was less than $\unit[10^{-9}]{\text{Torr}}$, it may still be possible for oxygen vacancies to be filled by oxygen species from the vacuum chamber. If such oxidation occurs, the conductance would drop even without the clustering effect. In order to distinguish the two effects, we carried out an experiment shown in Fig. \ref{oxidationSTO}. It shows how conductance changes in an oxygen-deficient STO surface as the sample is exposed to increasing oxygen pressure over a long period of time. 
Initially, the conductance showed a fast drop and then it changed to a slow decrease after about five hours. During that time, the partial oxygen pressure remained below $\unit[10^{-9}]{\text{Torr}}$. By increasing the oxygen pressure gradually to $\unit[510]{\text{Torr}}$ at around 50 hours, the conductance showed another fast drop followed by the same slow decrease as before. This slow background behavior did not change even after pumping out the oxygen down to $\unit[10^{-9}]{\text{Torr}}$ level after 120 hours. This clearly implies that there are at least two different mechanisms that can reduce the conductance. The sudden conductance drop, seen after introducing oxygen at around 50 hours, is clearly due to oxygen filling up the vacancies. However, the overall slow conductance decrease regardless of the background oxygen pressure cannot be explained by oxidation, and the only known mechanism that can explain such a phenomenon is the vacancy clustering.

Because clusters are energetically favorable \cite{luo:04,Cuong:07}, diffusion will force oxygen vacancies to migrate to form clusters, localizing otherwise itinerant electrons around them and causing the conductance to drop. The amount of mobile single vacancies will decrease as the clustering develops, and the clustering rate, thus conductance drop, will be high at the beginning and will reduce as time goes by. Therefore, the initial fast drop before 5 hours in Fig. \ref{oxidationSTO} could also be explained by clustering even if oxygen filling of the vacancies were completely forbidden. Although it is hard to separate out the vacancy-filling contribution to the time-dependent conductance drop, the important message is that vacancy clustering, which reduces the amount of mobile carriers for each oxygen vacancy, occurs actively in STO even at room temperatures.

\section{Conclusion}
In conclusion, the ion-milling process, a common dry-etching technique for complex oxide samples, creates on \STO\ surface a highly conducting layer much deeper than argon penerates, not only near room temperature but also at cryogenic temperatures where thermal diffusion is completely suppressed. At cryogenic temperatures, the conductance gradually grew and saturated to a highly metallic value, as a steady-state oxygen deficient layer developed within  \emph{hot zone} created by bombardment of the argon ions. Near room temperature, however, the condutance kept increasing without saturation because themal diffusion allowed oxygen vacancies to diffuse continuously into the bulk. These metal-insulator transitions driven by ion-bombardment process were well described by simple models assuming each oxygen vacancy donating two mobile electron carriers. The discrepancy between the measurements and the models can be accounted for by vacancy clustering, which makes oxygen vacancies less effective in donating electric carriers. Whether ion-bombardment-driven metal-insulator-transitions exist in materials other than \STO\ or not is an interesting open question for future studies. 

\begin{acknowledgments}
This work is supported by IAMDN of Rutgers University, National Science Foundation (NSF DMR-0845464) and Office of Naval Research (ONR N000140910749).
H. G. is also supported by the German academic exchange service (DAAD 221-ISAP D/07/16496).
\end{acknowledgments}


\end{document}